\documentclass[aps,darft,reprint,amsmath,amssymb,superscriptaddress]{revtex4-1}
\usepackage{graphicx}
\usepackage{color}

\usepackage{ifpdf}

\ifpdf
  \usepackage[pdftex]{graphicx}
  \DeclareGraphicsExtensions{.pdf}
  \usepackage[a4paper,dvipdfm,hyperindex=true]{hyperref}
\else
  \usepackage{graphicx}
  \DeclareGraphicsExtensions{.ps,.eps}
  \usepackage[a4paper,dvipdfm,hyperindex=true]{hyperref}
\fi

\definecolor{linkcol}{rgb}{0.2,0.2,0.6}

\hypersetup
{
bookmarksopen=true,
pdfmenubar=true, 
pdfhighlight=/O, 
colorlinks=true, 
pdfpagemode=None, 
pdffitwindow=true, 
linkcolor=linkcol, 
citecolor=linkcol, 
urlcolor=linkcol 
}

\usepackage{float}
\usepackage{subfigure}

\def\Tc{\ensuremath{T_{\rm c}}}
\def\Hc{\ensuremath{H_{\rm c}}}
\def\xc{\ensuremath{x_{\rm c}}}
\def\LiHoYF{LiHo$_x$Y$_{1-x}$F$_4$}

\def\cbl{\color{black}}

\begin{document}


\title{Phase diagram of diluted Ising ferromagnet \LiHoYF}

\author{P. Babkevich}
\email{peter.babkevich@gmail.com}
\affiliation{Laboratory for Quantum Magnetism, Institute of Physics, \'{E}cole Polytechnique F\'{e}d\'{e}rale de Lausanne (EPFL), CH-1015 Lausanne, Switzerland}

\author{N. Nikseresht}
 \affiliation{Laboratory for Quantum Magnetism, Institute of Physics, \'{E}cole Polytechnique F\'{e}d\'{e}rale de Lausanne (EPFL), CH-1015 Lausanne, Switzerland}

\author{I. Kovacevic}
 \affiliation{Laboratory for Quantum Magnetism, Institute of Physics, \'{E}cole Polytechnique F\'{e}d\'{e}rale de Lausanne (EPFL), CH-1015 Lausanne, Switzerland}

\author{J.~O.~Piatek}
 \affiliation{Laboratory for Quantum Magnetism, Institute of Physics, \'{E}cole Polytechnique F\'{e}d\'{e}rale de Lausanne (EPFL), CH-1015 Lausanne, Switzerland}

\author{B. Dalla~Piazza}
 \affiliation{Laboratory for Quantum Magnetism, Institute of Physics, \'{E}cole Polytechnique F\'{e}d\'{e}rale de Lausanne (EPFL), CH-1015 Lausanne, Switzerland}

\author{C.~Kraemer}%
 \affiliation{Laboratory for Quantum Magnetism, Institute of Physics, \'{E}cole Polytechnique F\'{e}d\'{e}rale de Lausanne (EPFL), CH-1015 Lausanne, Switzerland}

\author{K.~W.  Kr\"amer}
 \affiliation{Department of Chemistry and Biochemistry,  University of Bern, CH-3012 Bern,
Switzerland}

\author{K. Proke\v{s}}
 \affiliation{{Helmholtz-Zentrum Berlin f\"ur Materialien und Energie,  14109 Berlin Wannsee, Germany}}

\author{S. Mat'a\v{s}}
 \affiliation{{Helmholtz-Zentrum Berlin f\"ur Materialien und Energie,  14109 Berlin Wannsee, Germany}}

\author{J. Jensen}
 \affiliation{Niels Bohr Institute, Universitetsparken 5, 2100 Copenhagen, Denmark}

\author{H.M. R\o nnow }
 \affiliation{Laboratory for Quantum Magnetism, Institute of Physics, \'{E}cole Polytechnique F\'{e}d\'{e}rale de Lausanne (EPFL), CH-1015 Lausanne, Switzerland}

\date{\today}

\begin{abstract}
We present a systematic study of the phase diagram of \LiHoYF\ ($0.25 \leq x \leq 1$) Ising ferromagnets obtained from neutron scattering measurements and mean-field calculations. We show that while the  thermal phase transition decreases linearly with dilution, as predicted by mean-field theory, the critical transverse field at the quantum critical point is suppressed much faster. This behavior is related to competition between off-diagonal dipolar coupling and quantum fluctuations that are tuned by doping and applied field, respectively. In this paper, we quantify the deviation of the experimental results from mean-field predictions, with the aim that this analysis can be used in future theoretical efforts towards a quantitative description.
\end{abstract}

\maketitle

\section{\label{introduction} Introduction}
The Li$M$F$_4$ family, where $M$ is a rare-earth element, provides a series of materials for the study of magnetic dipolar interactions with negligible nearest- neighbor exchange coupling \cite{ AlsNielsen1974, Hansen1975, Misra1977, Magarino1980, Mennenga1984, Kraemer2012, babkevich-prl-2016}. Among them, the three-dimensional Ising ferromagnet LiHoF$_4$ has attracted much theoretical and experimental interest, offering a simple and well-understood Hamiltonian. The system is ferromagnetic below a Curie temperature of $\Tc\simeq1.53$\,K, and undergoes a quantum phase transition (QPT) when a field of 50\,kOe is applied transverse to the Ising axis \cite{Beauvillain1978, Christensen1979, Chakraborty2004, Ronnow2007}. The strong crystal field single-ion anisotropy at the Ho site results in an Ising groundstate doublet. However, unlike many other anisotropic dipolar-coupled systems where the same energy scale is relevant to the anisotropy and quantum fluctuations, \LiHoYF\ remains Ising-like well into the paramagnetic state above the critical transverse field (\Hc) \cite{Schechter2008}. At low temperatures, the ferromagnetic state is a result of competition between dipolar interactions which promote magnetic order and transverse magnetic field which introduces quantum fluctuations to destroy the ordered state. The role of the hyperfine interactions on the phase diagram close to the quantum critical region is discussed in Refs.~\onlinecite{Schechter2008, Ronnow2007, Ronnow2005}. Careful studies of the excitation spectra while tuning through the quantum critical point (QCP) have revealed that the hyperfine coupling to the nuclear spin bath forestalls the electronic mode softening expected for a quantum phase transition. As a result, the ferromagnetic-paramagnetic phase boundary at $T=0$ is strongly modified by the hyperfine interaction.

The ability to dilute Ho$^{3+}$ sites with non-magnetic Y$^{3+}$ ions provides a rich arena to explore how disorder and quantum fluctuations affect the magnetic properties and collective phenomena \cite{Schechter2008, Silevitch2010, Reich1990, Tabei2006}. The \LiHoYF\ system adopts the scheelite structure for all values of $x$ and allows a systematic study of dilution of magnetic moments from a correlated ferromagnet to a single-ion state. Ferromagnetic order persists with decreasing \Tc\ until a critical Ho concentration $\xc \approx 25\%$ is reached \cite{Alonso2010}. Below \xc, the induced local randomness is effective enough to destroy long-range order. The nature of the highly diluted systems for $x<0.25$ has been debated in a series of experimental and theoretical reports. Some studies have reported that there exists a transition into a spin-glass state, however numerical calculations were not able to find evidence for a spin-glass transition \cite{Wu1993, Tam2009, Rosenbaum1999, Reich1990, Biltmo2007, Biltmo2008}. The interplay between frustrated dipolar coupling, local disorder induced by chemical doping, and quantum fluctuations due to a transverse field enhance the complexity of the system, which make the correct interpretation important \cite{Schechter2007, Ghosh2002, Reich1987}.

In order to clarify the ambiguities of the very diluted phases, first one needs to properly understand the role of the above mentioned terms on the diluted series, where the long-range order is shown to survive \cite{Silevitch2007}. It has been argued that in \LiHoYF\ when $x\neq1$, due to the breaking of symmetry, transverse couplings between Ho$^{3+}$ moments no longer cancel out which leads to additional longitudinal and transverse internal random fields \cite{Schechter2008, Tabei2006}. Correction for these terms in the effective Hamiltonian leads to a modification of the ferromagnetic-paramagnetic transition in \LiHoYF\ ($x\neq1$). Even though considerable studies have been performed on \LiHoYF, especially on very diluted concentrations, a detailed comparison of a series of compounds is missing. In this work we systematically track the evolution of disorder by gradually doping the ferromagnetic systems with $0.25\leq x\leq 1$. For all concentrations the classical and quantum critical points are measured by means of neutron scattering. The ordered moment and critical scattering are traced as a function of $x$, and the entire phase diagrams are mapped. For theoretical comparison, mean-field calculations were performed, and the discrepancies with experimental data are quantified. The experimental dependence of \Hc\ on Ho$^{3+}$ concentration could be reproduced provided the ordered moment is reduced.

The outline of the paper is as follows: In Sec.~\ref{Phase diagram studies by MF calculations} a basic introduction to mean-field calculations of \LiHoYF\ is given. Neutron scattering measurements as a function of temperature and magnetic field are discussed in Sec.~\ref{Neutron diffraction} and results are compared to simulations. The differences between measured and theoretical results are analyzed in Sec.~\ref{Discussion} and the probable reasons for the observed discrepancies are discussed.

\section{\label{Phase diagram studies by MF calculations} Phase diagram studies by mean-field calculations}

The full rare-earth Hamiltonian was diagonalized within the virtual-crystal mean-field (VCMF) approximation, widely used to solve the mean-field Hamiltonian of composites. The magnetic moment operators of Ho$^{3+}$ and Y$^{3+}$ ions at site $i$ are labeled as
\begin{equation}
\mathbf{J}_i=x\mathbf{J}_i^{\rm Ho}\oplus(1-x)\mathbf{J}_i^{\rm Y}.
\label{eqn:Jop}
\end{equation}
The virtual crystal method uses a homogeneous approximation with a composite moment where $x$ is the proportion of Ho$^{3+}$ ions. Although in the present case the dilution is made by non-magnetic Y$^{3+}$, this approach can be equally well applied for a mixture of magnetic ions, e.g., Ho$^{3+}$ and Er$^{3+}$ \cite{Piatek2013}. The numerical algorithm is essentially a simple mean-field calculation, with the difference that VCMF not only mixes site-specific mean-fields, but also the ion type specific mean-fields \cite{Kraemer2012}. If there exists only one kind of rare-earth ion in the system, VCMF is equivalent to a mean-field treatment \cite{Ronnow2007}.

The complete Hamiltonian of the system includes five terms,
\begin{eqnarray}
\mathcal{H} &=&\mathcal{H}_{\rm CF}+\mathcal{H}_{\rm hyp}+\mathcal{H}_{\rm Z}+\mathcal{H}_{\rm D}+\mathcal{H}_{\rm ex} \nonumber \\
&=&\sum_i \left[\mathcal{H}_{\rm CF}(\mathbf{J}_i) + A \mathbf{J}_i \cdot \mathbf{I}_i- g\mu_B \mathbf{J}_i.\mathbf{H} \right] \nonumber \\
& & -\frac{1}{2}\sum_{ij} \sum_{\alpha\beta}\mathcal{J}_{\rm D} \overline{\overline{D}}_{\alpha\beta} \mathbf{J}_{i\alpha} \mathbf{J}_{j\beta} - \frac{1}{2}\sum_{\langle ij\rangle}\mathcal{J}_{\rm ex}\mathbf{J}_i \cdot \mathbf{J}_j \,
\end{eqnarray}
where $\mathcal{H}_{\rm CF}$ is the crystal field, $\mathcal{H}_{\rm hyp}$ is the hyperfine coupling of Ho$^{3+}$ ion to its $I=7/2$ nuclear spin, $\mathcal{H}_{\rm Z}$ the Zeeman effect, $\mathcal{H}_{\rm D}$ the dominant dipolar interaction, and $\mathcal{H}_{\rm ex}$ the nearest neighbors Heisenberg exchange interaction. Owing to spin-orbit interaction, $J=8$ in the groundstate. The hyperfine coupling constant of Ho$^{3+}$ is taken to be $A=3.36$\,$\mu$eV as reported from hyperfine resonance and specific heat measurements \cite{Magarino1980, Kjaer1999, kovacevic-submitted}.

The crystal-field parameters used in this study are those reported in Ref.~\onlinecite{Ronnow2007}. We assume that the crystal field surrounding a Ho$^{3+}$ ion is not perturbed significantly by dilution \cite{babkevich-prb-2015}.
The crystal field splits the $J=8$ ground state of Ho$^{3+}$ for point symmetry $\bar{4}$ into four doublet and nine singlet crystal-field levels, where the lowest crystal-field level is a doublet. Its wavefunction is a linear combination of $m_J = \pm 7, \mp 5, \pm 3, \mp 1$. The dominance of high $m_J$ contributions give it an Ising anisotropy.
The next crystal-field level is found at about 11\,K. The strength of the superexchange coupling, $\mathcal{J}_{\rm ex}$, was fixed to $-0.1\,\mu$eV in our calculation.
%
Furthermore, to account for the strong $z$-axis fluctuations established by the $1/z$ expansion \cite{Ronnow2007}, the mean-field, i.e., the ferromagnetic interaction times the local magnetic moment, is scaled by the factor 0.785 in combination with a minute adjustment of the crystal field. The results derived in this \emph{effective MF approximation} are found to agree accurately with the predictions of the first-order theory in $1/z$ at $x=1$, and also,  the effective VCMF in the dilute case accounts in an acceptable way for that obtained when combining the $1/z$ expansion with the coherent potential approximation (CPA) (see Ref.~\onlinecite{jensen-jopc-1984}). Within the effective MF approximation, the calculated \Tc\ is higher than found experimentally for LiHoF$_4$ of $\Tc=1.53$\,K.
{\cbl Indeed, the phase boundary close to \Tc\ cannot be accurately reproduced theoretically and remains an outstanding problem \cite{Chakraborty2004, Tabei2008, dunn-prb-2012}.}
The long-range nature of dipole-dipole interactions was treated by splitting the dipolar fields' summation into a short-range discrete sum over 50 unit cells and a continuous integration towards the sample boundaries assuming a spherical shape of the sample \cite{Tabei2008, Jensen1991}.

The homogeneous distribution of the ions within the lattice implies that off-diagonal interactions cancel out due to symmetry. Therefore, local randomness and frustration will not be accounted for in this picture. As will be discussed later, these absent terms play an important role on the critical properties of the diluted systems and neglecting them results in discrepancies with experimental observations.

\section{\label{Neutron diffraction} Neutron diffraction}

Neutron scattering measurements were performed on single-crystals of \LiHoYF, grown by the Bridgman technique with nominal Ho$^{3+}$ concentrations of $x= 0.25$, 0.33, 0.46 , 0.67, and 1. In order to reduce the neutron absorption, all compounds were synthesized from isotopically enriched $^7$Li ($>$99.9\%). The Ho$^{3+}$/Y$^{3+}$ ratio was checked by means of energy-dispersive x-ray spectroscopy, and the nominal Ho$^{3+}$ concentration was confirmed to be accurate to within $\pm$1\% through density measurements of the crystal with a differential air-liquid weighing technique. The density effect of the utilized liquids was negligible since similar results were obtained using water, ethanol, and isopropanol. The surface tension of the liquid was effectively reduced by using thin copper wires with a diameter of 50\,$\mu$m to carry the samples. The crystal structure of \LiHoYF\ crystallizes in space group $I4_1/a$. The lattice parameters slightly vary with $x$ and are approximately $a \simeq 5.18$\,\AA\ and $c \simeq 10.75$\,\AA.

The experiments on crystals with $x=0.25$, 0.46, 0.83, and 1  were performed on the E4 two-axis diffractometer at the Helmholtz-Zentrum Berlin. The crystals with $x=0.33$ and 0.67 were measured on the RITA-II triple-axis spectrometer at the Paul Scherrer Institut and the D23 and D10 diffractometers at the Institut Laue Langevin, Grenoble.

To improve the thermalization of the samples below 1\,K, crystals were glued with Stycast onto sample holders made of oxygen-free copper. The samples were mounted inside a dilution refrigerator and vertical field superconducting magnets. Scans were performed in the $(h0l)$ scattering plane such that the magnetic field ($H$) was applied along the crystallographic $b$-axis. Scans were centered on the $\mathbf{Q}=(200)$ reflection, and the wavevector of the incoming neutron beam was selected in the 2.56--2.66\,\AA$^{-1}$ window in different series of the measurements.

Due to ferromagnetic order in the \LiHoYF\ systems studied, for a given reflection the total neutron scattering cross-section is composed of both nuclear and magnetic contributions. In this paper, we shall focus on the intensity due to magnetic scattering, i.e., $I-I_{\rm N}$, where $I$ is the total intensity and $I_{\rm N}$ is intensity from nuclear scattering which we obtain from measurements above \Tc. Since neutron scattering from ferromagnetically ordered compounds is proportional to the square of the ordered moment perpendicular to $\mathbf{Q}$, one can extract the phase transition temperature (or equivalently the field) at which the compounds order magnetically. Resolution-limited Bragg peaks at base temperature in the ferromagnetic phase indicate long-range order in all samples. To fit the crystal rotation $\omega$ scans we have used,
\begin{equation}
{I}(\omega) \propto [A\delta(\omega - \omega_0) + BL(\omega-\omega_0)]\otimes G(\omega),
\label{eqn:bp_cs}
\end{equation}
where $A$ is the scattering intensity from long-range order, centered at $\omega=\omega_0$; in our scattering geometry and zero applied field $A\propto M_c^2$, where $M_c$ is the magnetization along $c$. In our measurements $A\gg B$, where $B$ denotes the amplitude of critical scattering described by a Lorentzian $L$. A convolution with a Gaussian $G$ is made to account for the instrumental resolution.

\subsection{Temperature dependence}
\label{sec:temperaturedep}

\begin{figure}
\includegraphics[width=\columnwidth]{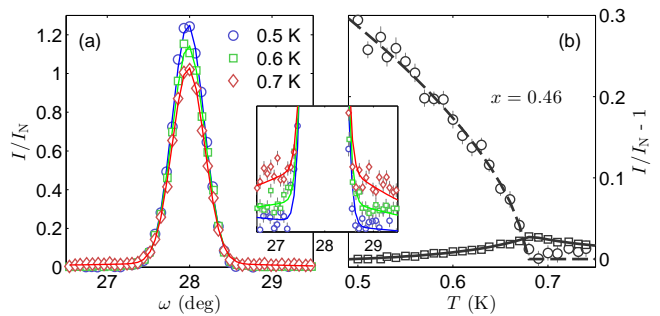}
\caption{\label{46T} (color online) (a) Crystal rotation scans through $\mathbf{Q}=(200)$ for the $x=0.46$ at three different temperatures. The red lines are fits described in the text. The inset demonstrates the enhancement of critical scattering around $\Tc(x=0.46)\approx 0.68$\,K. (b) Scattering from long-range magnetic order (circles) and the critical scattering (squares) as a function of temperature. The lines are the power-law fits to the data.}
\end{figure}

Figure~\ref{46T}(a) shows transverse scans through the $\mathbf{Q}=(200)$ reflection in the $x=0.46$ sample at three different temperatures close to $\Tc(x=0.46)\approx 0.68$\,K. In this compound, nuclear scattering dominates $(200)$ reflection intensity. Scans at each point in temperature were extended in reciprocal space such that the diffuse component as well as the Bragg scattering were covered. Figure~\ref{46T}(b) shows the onset of long-range magnetic order close to $\Tc(x=0.46)$ as indicated by the increase in the Bragg peak intensity. Critical scattering is found to be strongest at the phase transition as expected. This data is representative of the salient features found from neutron scattering in the other samples.

\begin{figure}
\includegraphics[width=\columnwidth]{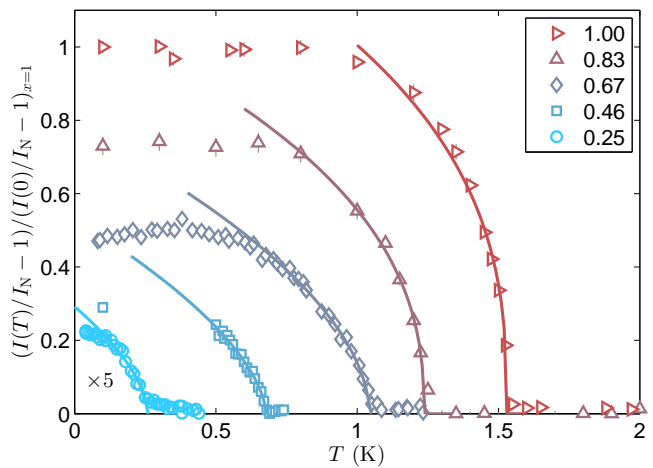}
\caption{\label{I-T} (color online) Temperature dependence of the magnetic intensity of \LiHoYF\ for $0.25\leq x\leq1$. To ease comparison, for each compound we normalize the ratio of magnetic to nuclear intensity by the ratio of magnetic to nuclear intensity of LiHoF$_4$ at base temperature. Solid lines show power-law fits from which the critical temperature was extracted as described in the text.}
\end{figure}

Figure~\ref{I-T} shows the temperature dependence of the magnetic scattering at the (200) Bragg point for increasing Y$^{3+}$ content in \LiHoYF. For better comparison, for all compounds the contribution of nuclear scattering $I_{\rm N}$ has been subtracted from the total intensity $I(T)$ using measurements in the paramagnetic state. We consider the ratio of the magnetic to nuclear scattering [$(I(T) - I_{\rm N})/I_{\rm N}$] which is normalized by the magnetic scattering cross-section of the parent compound LiHoF$_4$ at base temperature [$(I(0) - I_{\rm N})/I_{\rm N}$]. With increasing Y content \Tc\ shifts to lower temperatures. At the same time, the total strength of scattering is also reduced -- as expected since magnetic Ho$^{3+}$ ions are replaced by non-magnetic Y$^{3+}$ ions. For the $x=0.67$ compound, largest intensity is observed above base temperature. The origin of this effect is not clear. We note that weak scattering from critical fluctuations is also observed, as already discussed for $x=0.46$. The line-shape of the critical scattering seems comparable for the $x=0.25$, 0.46, and 0.67 samples, as well as the ratio between the base temperature intensity and the maximum intensity of the critical scattering at the transition point. Within the instrumental resolution, we did not find a significant contribution from critical scattering for samples with $x=0.83$ and 1.

\begin{figure}
\includegraphics[width=\columnwidth]{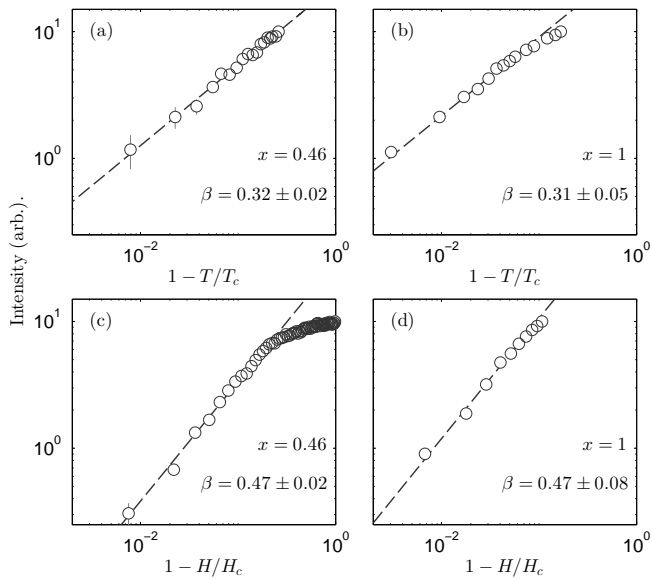}
\caption{\label{fig:critical} (color online) Intensity of the magnetic Bragg peak $(200)$ close to the phase transition in \LiHoYF\ for $x=0.46$ and $1$. The critical exponent $\beta$ is shown for each scan and composition.}
\end{figure}

\begin{figure}
\includegraphics[width=\columnwidth]{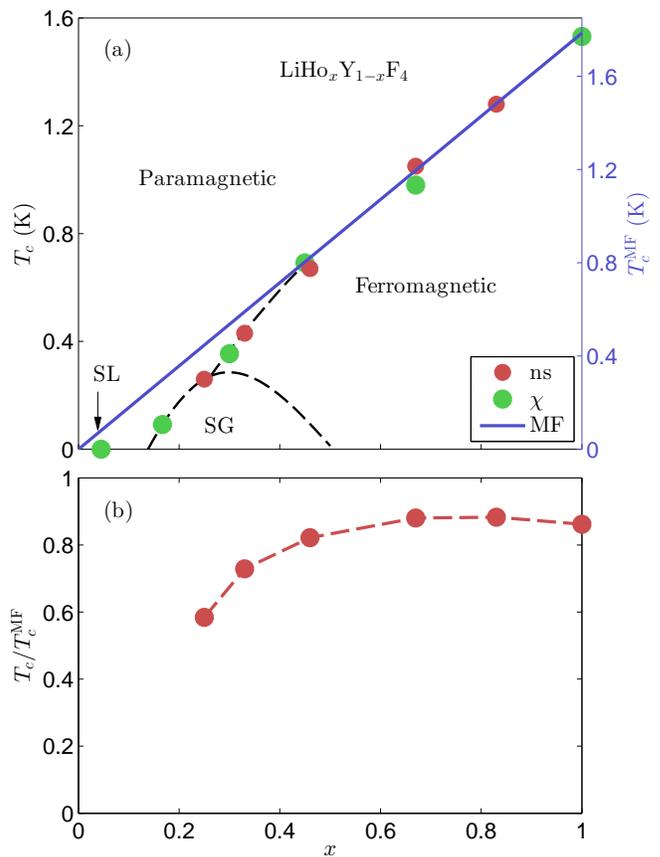}
\caption{\label{Tc-x} (color online) (a) The magnetic phase diagram of \LiHoYF\ as a function of Ho$^{3+}$ content $x$. The Curie temperature \Tc\ from neutron scattering (ns) is compared with magnetic susceptibility data ($\chi$) from  Ref.~\onlinecite{Reich1990}. The blue line denotes the mean-field prediction. At low Ho$^{3+}$ concentrations the emergence of spin-glass (SG) and spin-liquid (SL) phases was reported. (b) Extracted \Tc\ for each compound normalized to its corresponding mean-field value. The dashed line is a guide to the eye.}
\end{figure}

{\cbl
\begin{table}
  \centering
  \caption{Extracted critical temperature and field as a function of doping in \LiHoYF. The uncertainty of the values is shown in parentheses.}
  \begin{tabular}{lccccc}
    \hline
    \hline
     $x$ & 0.25 & 0.46 & 0.67 & 0.83 & 1.00 \\
     \hline
    \Tc\ (K)   & 0.25(1) & 0.677(4) & 1.04(3) & 1.24(1) & 1.530(2) \\
    \Hc\ (kOe) & 5.73(8) & 11.9(1) & 21.5(6) & 36.0(6) & 47.3(1) \\
    \hline
    \hline
  \end{tabular}
  \label{tab:HcTc}
\end{table}
}

%

{\cbl
In order to extract a value of \Tc\ for each compound we employed power-law fits in the vicinity of \Tc\ using the standard definition in the fitting, where $I\propto |1-T/\Tc|^{2\beta}$. The values of \Tc\ found for each compound are listed in Table~\ref{tab:HcTc}. In Figs.~\ref{fig:critical}(a) and \ref{fig:critical}(b) we show the power-law temperature dependence of Bragg peaks for the $x=0.46$ and $x=1$ compounds. In both cases we find a value of $\beta_{\rm T} = 0.32(5)$ that corresponds to the renormalization group result of $\beta_{\rm RG} = 1/3$ for the three-dimensional Ising model.
}

Figure~\ref{Tc-x}(a) shows the results of our measurements. The extracted critical temperatures are in good agreement with the susceptibility results reported previously in Ref.~\onlinecite{Reich1990}. For high Ho$^{3+}$ concentrations of $x\gtrsim0.5$, linear suppression of \Tc\ by doping Y$^{3+}$ is observed in agreement with the predictions of mean-field theory $\Tc(x)=x \cdot \Tc(x=1)$. A comparison of the experimental and measured values of \Tc\ obtained is shown in Fig.~\ref{Tc-x}(b). When the Ho$^{3+}$ concentration is lowered below $x \approx 0.4$, \Tc\ deviates significantly from mean-field approximation. This is shown experimentally in Fig.~\ref{Tc-x}(a) to be related to the onset of the spin-glass phase. Below a marginal concentration around $x \approx 0.2$ ferromagnetic ordering is circumvented by the interplay of disorder and frustration arising from the anisotropy of dipolar interaction. On further dilution the cross over temperature decreases to zero and for concentration around $x \approx 0.05$ there is no spin freezing at all. This phase is referred to as spin liquid anti-glass or decoupled cluster glass due to its unusual behavior \cite{Ghosh2002}.

\subsection{Transverse field dependence}

To study the effect of the transverse magnetic field, neutron scattering measurements were performed with field perpendicular to the scattering plane, along the crystallographic $b$-axis. In the zero field cooled (ZFC) measurements, where the samples were cooled in zero field down to base temperature ($T\approx100$\,mK), the resolution-limited magnetic Bragg peak revealed the presence of the long-range order in all compounds, as discussed in Sec.~\ref{sec:temperaturedep}. However, below around 67\% Ho$^{3+}$ concentration a path dependence was observed depending on the annealing protocol and long-range order was entirely suppressed in the $x=0.25$ sample in field cooled studies.\cite{Kraemer2009} We will address these interesting results elsewhere and only focus on ZFC measurements here.

\begin{figure}
\includegraphics[width=\columnwidth]{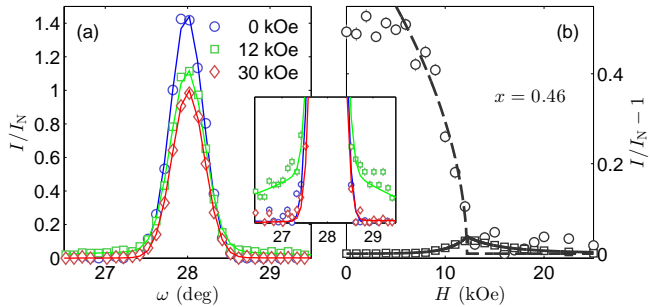}
\caption{\label{46H} (color online) (a) Crystal rotation scans through $\mathbf{Q}=(200)$ for the $x=0.46$ sample at three different fields. The solid lines are fits as described in the text. The critical scattering is strongest around \Hc\ (inset). (b) Field dependence of the magnetic Bragg peak intensity (circles) and critical scattering (squares). The plotted lines are power-law fits to the data.}
\end{figure}

Crystal rotation scans through the $\mathbf{Q}=(200)$ magnetic reflection were performed to simultaneously investigate the order parameter and diffuse scattering as a function of transverse field. Figure~\ref{46H}(a) shows typical $\omega$ rotation scans on $x=0.46$ concentration at three different magnetic fields: below, at, and above the critical field. As can be observed at the tails of the Bragg peak [see Fig.~\ref{46H}(a) inset], critical scattering is enhanced close to \Hc. The analysis of the scans was treated in in the same way as zero field temperature dependence studies. The intensity of the Bragg peak is proportional to the order parameter squared, and the critical scattering extracted using Eq.~\ref{eqn:bp_cs} is shown in Fig.~\ref{46H}(b).

\begin{figure}
\includegraphics[width=\columnwidth]{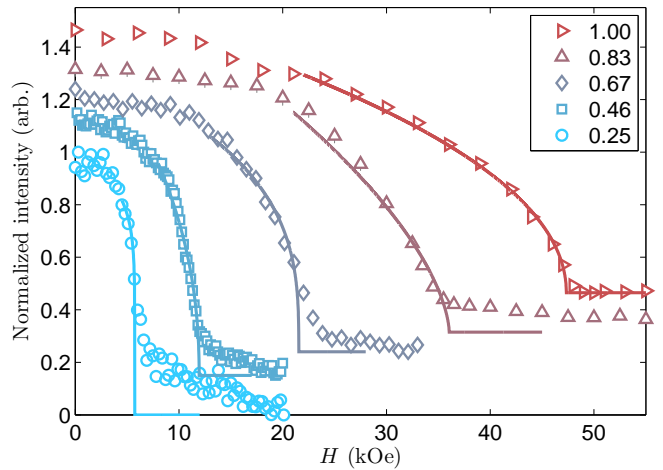}
\caption{\label{I-H} (color online) Bragg peak intensity as a function of $H$ in \LiHoYF\ at 100\,mK. To ease the comparison, the curves are normalized to the maximum and minimum counts measured. The solid lines are power-law fits to the Bragg peak. The data has been displaced vertically. }
\end{figure}

The normalized Bragg peak intensity of the measured compounds as a function of $H$ is shown in Fig.~\ref{I-H}. We can accurately extract the critical fields at different Ho$^{3+}$ concentrations by power-law fits to the intensity close to \Hc. In the pure LiHoF$_4$, the intensity of the (200) reflection gradually decreases, with magnetic contribution vanishing at 47.3(1)\,kOe. Above 50\,kOe, we again observe an increase of scattering by about 10\% at 100\,kOe which we attribute to the magnetization of Ho$^{3+}$ moments along the field direction which is out of the scattering plane. This effect is also captured in our mean-field calculations. We find no sign of critical scattering within the instrument resolution. In the case of $x=0.83$ there is a steady decrease of intensity above $\Hc(x=0.83)\approx 36$\,kOe. Surprisingly, in our measurements the (200) reflection at base temperature and 80\,kOe has less intensity than above $\Tc(x=0.83)$ at 2\,K and in zero field. That is, there is less scattering in the quantum paramagnetic state than from the crystal lattice by approximately 20\%. The origin of this is unclear to us but may be related to a structural distortion or a straining of the crystal under a magnetic field leading to a change in the extinction contribution. The measurements on $x=0.67$, 0.46, and 0.25 are qualitatively similar with decreasing \Hc\ on decreasing $x$.
The evolution of the order parameter as a function of temperature (Fig.~\ref{I-T}) and magnetic field (Fig.~\ref{I-H}) near the critical points differ. The thermal transition is well-defined in Fig.~\ref{I-T}. However, a long tail above \Hc\ (for $x<1$) is observed. This could be due to a difference of critical fluctuations close to \Hc\ and \Tc; or possibly a signature of a crossover into a quasi-spin-glass phase predicted by numerical calculations \cite{andresen-prl-2013}.
%
%
{\cbl
A summary of the extracted values of \Hc\ are listed in Table~\ref{tab:HcTc}. In Figs.~\ref{fig:critical}(c) and \ref{fig:critical}(d), we show that $\beta_{\rm H} = 0.47(8)$, which is close to the mean-field exponent of 0.5. Within the uncertainty of the fits, the critical exponent does not change on dilution.
}

\subsection{Magnetic phase diagram}

\begin{figure}
\includegraphics[width=0.95\columnwidth]{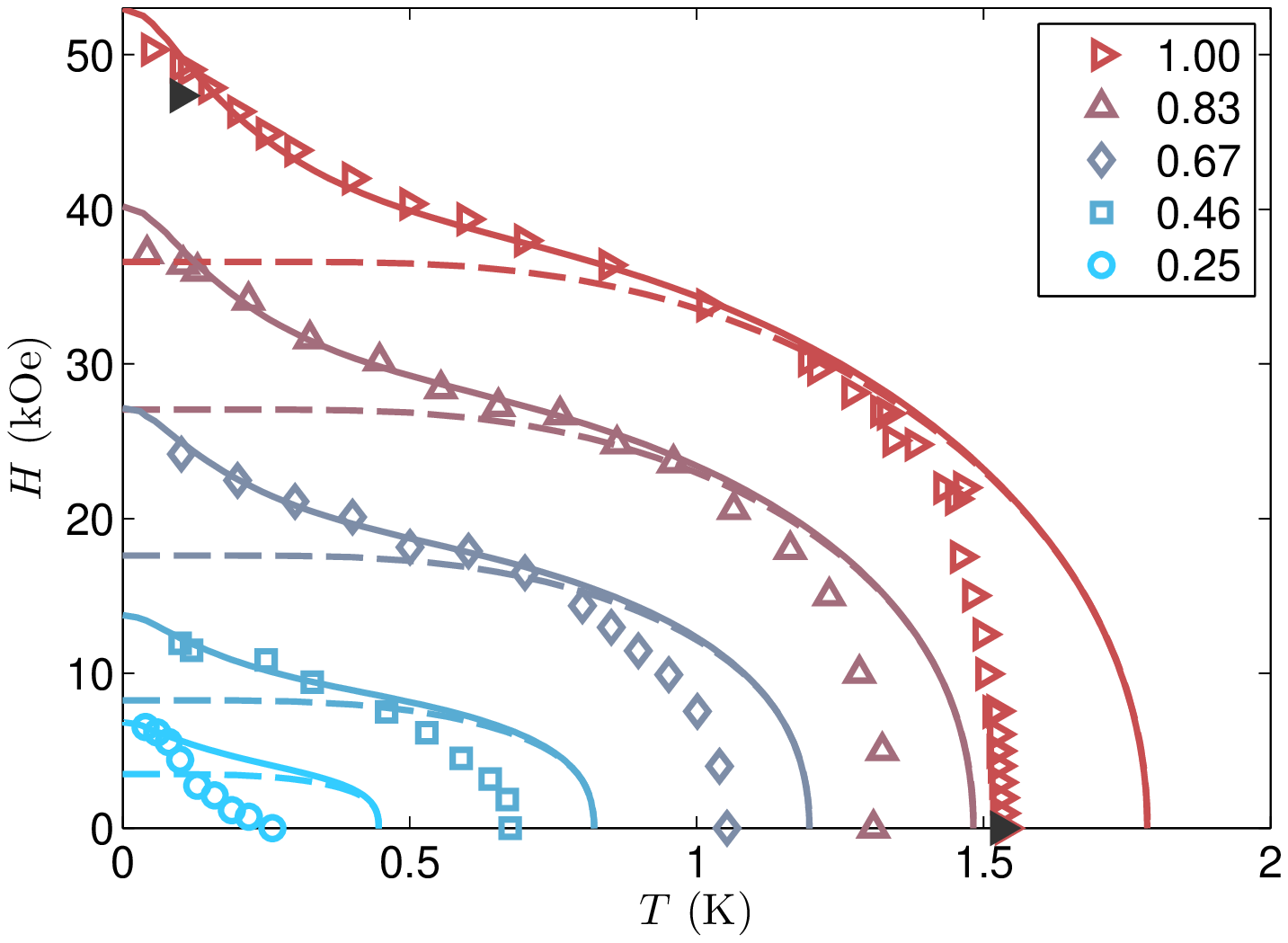}
\caption{\label{T-H} (color online) Phase diagram of \LiHoYF\ obtained from neutron diffraction (symbols).
{\cbl
Data for $x=1$ are from ac-susceptibility, thermal expansion, and magnetostriction measurements in Refs.~\onlinecite{Bitko1996, dunn-prb-2012}; the values of \Tc\ and \Hc\ obtained from neutron scattering measurements on $x=1$ are plotted by solid black triangles. }
The solid (dashed) lines are mean-field calculations with (without) the hyperfine interaction. The mean-field curves are scaled (see Fig.~\ref{H-x}) to match the low-temperature range of the phase diagram.}
\end{figure}

Neutron scattering allows us to accurately extract the temperature and field dependence of the order parameter. The quenching of ferromagnetic order, is indicated by the disappearance of magnetic scattering. By these means we have mapped out the temperature-field phase diagram of \LiHoYF\ for $x=0.25$, 0.46,  0.67, and 0.83 as shown in Fig.~\ref{T-H}. The data for the $x=1$ concentration were extracted from ac-susceptibility measurements reported by Bitko~\emph{et al.},\cite{Bitko1996}.
{\cbl Our measurements of the magnetic phase boundaries are in excellent agreement with those previously obtained from ac-susceptibility measurements for $x=0.44$ and 0.65 \cite{Silevitch2007}. }
Our data can be readily compared to virtual-crystal mean-field calculations. The field values have been scaled to match the low temperature range of the experimental data. Doing this improves the phase boundary comparison, and helps the zero temperature \Hc\ extrapolation. The scaling factor increases from 1 for the pure compound to 4 for the $x=0.25$ sample, which shows the enhancement of the discrepancy from mean-field by the introduction of frustration and disorder into the system. In all systems studied, we find that the critical field is enhanced below approximately $T \approx \Tc/2$. This enhancement is well captured in our calculations by the introduction of hyperfine coupling. As has already been discussed in Sec.~\ref{sec:temperaturedep}, mean-field calculations do not agree very well with experiment close to \Tc\ predominantly due to the absence of fluctuations in our model [see Fig.~\ref{Tc-x}(b)].

\begin{figure}
\includegraphics[width=0.95\columnwidth]{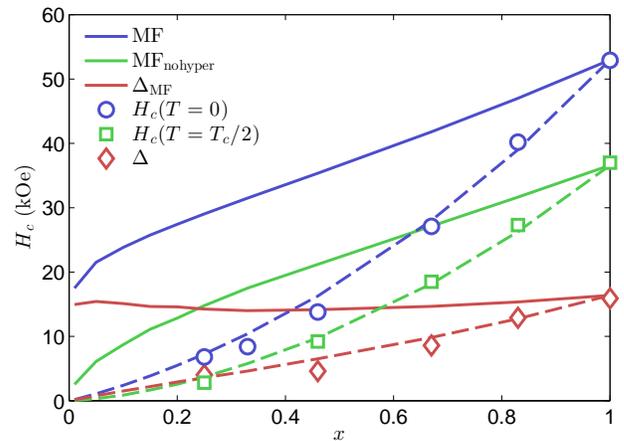}
\caption{\label{H-x} (color online) Critical field behavior as a function of Ho$^{3+}$ content $x$ in \LiHoYF\ obtained from neutron scattering (blue circles). The fields at $\Tc/2$ represent hyperfine-free values of \Hc\ (green squares). The difference, $\Delta$, between these two values demonstrates the hyperfine effect (red diamonds). The solid lines are mean-field calculations in the presence (absence) [blue (green) solid line] of hyperfine interactions, and the solid red line is the subtraction of the two curves. The dashed lines are \Hc\ values which are obtained by rescaling the critical field by $x$.}
\end{figure}

Figure~\ref{H-x} shows the dilution dependence of the critical field in \LiHoYF. We consider three values obtained experimentally: (i) critical field $\Hc(T=0)$ extrapolated to $T=0$, (ii) $\Hc(T=\Tc/2)$ which represents the critical field in the absence of hyperfine interaction, and (iii) the difference between (i) and (ii). From the phase diagrams in Fig.~\ref{T-H}, we expect that in the absence of the hyperfine interaction, the critical field for a given dilution does not change appreciably below $\Tc/2$. Therefore, the corresponding $H$ at this temperature can be considered to be the hyperfine-free value of \Hc\ for each sample. We find from our measurements that $\Hc(0)$ decreases almost quadratically from 53\,kOe in $x=1$ to 7\,kOe at 25\% Ho$^{3+}$ concentration. The hyperfine-free critical field $\Hc(\Tc/2)$ extracted from our data and the difference $\Hc(0)-\Hc(\Tc/2)$ follow a similar trend.

In comparing our experimental results with VCMF calculations, we find that $\Hc(0)$ and $\Hc(\Tc/2)$ should increase almost linearly for $0.2 < x < 1$. The difference $\Hc(0)-\Hc(\Tc/2)$, which is attributed to electro-nuclear coupling, reveals that hyperfine contribution on \Hc\ is nearly dilution independent. The effect of dilution on the criticality around the quantum critical point is poorly captured in this model. Therefore, our model gives very different results to what we have observed. The experimental results imply that the system is more easily disordered by an external magnetic field than expected.

Phenomenologically we can rescale the critical field found from VCMF by dilution, i.e., $\Hc \rightarrow x \Hc$. By doing this we reach a good agreement between VCMF and measured results over the entire dilution range studied. An alternate strategy would be to consider a model where the mean-field moment within VCMF is reduced by $x^2$ rather than $x$, i.e., Eq.~\ref{eqn:Jop} becomes $\mathbf{J}_i = x^2\mathbf{J}_i^{\rm Ho}$. Since we are reducing the effective moment of Ho$^{3+}$ ions, this would also dramatically affect the expected \Tc.

\begin{figure}
\includegraphics[width=\columnwidth]{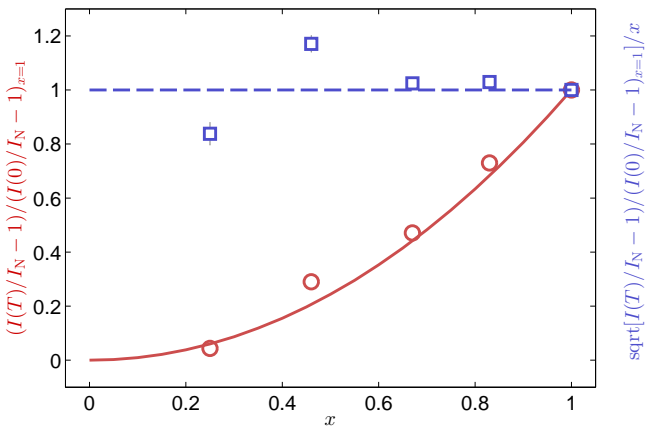}
\caption{\label{orderedmoment-x} (color online) The ratio of the magnetic to nuclear intensities as a function of Ho$^{3+}$ content $x$ in \LiHoYF\ at base temperature and $H=0$ (red circles), normalized to the value for $x=1$. The red solid line is the calculated ratio of the neutron scattering cross-sections. The effective magnetic moment per Ho$^{3+}$ ion is plotted as blue squares with a dashed line serving as a guide to the eye.}
\end{figure}

To determine whether it is necessary to scale the ordered moment in VCMF by $x^2$ we turn our attention to the size of magnetic moment on Ho$^{3+}$ ions obtained in our neutron diffraction measurements. The ratio between the magnetic ($I-I_{\rm N}$) and nuclear ($I_{\rm N}$) intensities at base temperature and zero field decrease approximately with dilution squared, shown in Fig.~\ref{orderedmoment-x}. We find that the ratio of the magnetic to nuclear scattering can be well described by our calculations of neutron scattering cross-section where the electronic moments scale as $x\langle J_z \rangle$. Since the scattering lengths of Y$^{3+}$ (7.75\,b) and Ho$^{3+}$ (8.01\,b) are very similar, the nuclear structure factor does not have a significant dependence on dilution. The strength of magnetic scattering is proportional to the square of the ordered moment. Therefore, the ordered moment relative to that of Ho$^{3+}$ ions in LiHoF$_4$ can be extracted for each concentration, as shown in Fig.~\ref{orderedmoment-x}. Within the accuracy of our measurements, the ordered moment per Ho$^{3+}$ ion in fact remains unchanged for the concentrations measured.

\section{Discussion}
\label{Discussion}

The competition between random local magnetic fields through substitution of magnetic Ho$^{3+}$ ions with nonmagnetic Y$^{3+}$ in \LiHoYF\ and quantum fluctuations from the transverse magnetic field yield a complex plethora of physics.
{\cbl
At base temperature the measured critical field decreases much faster than the mean-field prediction, but we have found that a very simple phenomenological rescaling by $x$ of the calculated critical field yields very good agreement with the measured phase boundary. Our results show that the mean-field calculation overestimates \Tc\ by 15-20\% down to $x=0.5$, below which the measured \Tc\ decreases at a faster rate [see Fig.~\ref{Tc-x}(b)].
}
%
%
%
%
%
By scaling the mean-field, our MF model accounts effectively for that part of the thermal and random-field fluctuations which may derive from an effective averaged medium (as assumed to first order in the $1/z$ expansion when combined with the CPA). The non-linear variation of \Tc\ indicates an additional complex interplay of fluctuations and randomness in  \LiHoYF\ close to \Tc\ which requires further theoretical studies.
Moreover, for the lowest concentrations of Ho$^{3+}$ in \LiHoYF, the system may be entering into a dipolar quasi-spin-glass state as we increase $H$. The long tail above \Hc\ in the order parameter curves of systems with $x<0.7$, as shown in Fig.~ \ref{I-H}, is possibly a signature of such a cross-over when the temperature is sufficiently low \cite{andresen-prl-2013}.

Our experimental results shown in  Fig.~\ref{H-x}, demonstrate the increasingly stronger reduction of \Hc\ as the concentration of Ho$^{3+}$ is reduced compared to our VCMF model. Random fields together with quantum fluctuations are at the heart of this phenomenon \cite{sch2006, Tabei2006, Tabei2008, schechter-prb-2008}. Numerical calculations for $x = 0.5$ and 0.7 found that \Hc\ is significantly suppressed when off-diagonal bilinear dipolar terms are included in the effective Hamiltonian \cite{Tabei2006}. The rate of suppression of \Hc\ is greater for lower concentrations of Ho$^{3+}$ \cite{Tabei2006}, which is also in agreement with our results. However, these calculations do not include the hyperfine interactions, which plays an important role in stabilizing the magnetic order. At lowest values of $x$, when the effective dipolar interactions are weak, the phase boundaries are mostly dictated by hyperfine interaction. This feature is obvious in the $x=0.25$ phase diagram, where the effect of the hyperfine interactions is more pronounced. Similar features have been argued by Schechter~\emph{et al.} \cite{Schechter2008}, in comparing two highly diluted systems, where a re-entrance and non-monotonic $H$ suppression with $x$ was predicted. We hope that the experimental account of the field, temperature, and dilution dependence reported herein will inspire a detailed theoretical description taking into account both electronic and nuclear degrees of freedom as well as random longitudinal and transverse fields.

\section{Conclusion}

We have analyzed the result of the competition between long-range order and random fields on the criticality of the ferromagnetic \LiHoYF\ with $0.25\leq x\leq1$ using neutron scattering and mean-field calculations. Transverse-dipolar interactions are responsible for the deviation of the classical phase transition from mean-field prediction below $x=0.5$. The quantum phase transition is found to be more sharply suppressed by dilution than expected from mean-field calculations that we account for by a phenomenological rescaling of \Hc\ by $x$. At base temperature and low magnetic fields where the energy scale of electron and nuclear spin coupling is effectively higher than the dipolar interactions, magnetic order of the Ho$^{3+}$ ions is observed for all studied compositions $x$. However, the persisting long tails of scattering intensity above \Hc\ could indicate the appearance of a quasi-spin-glass phase. Off-diagonal coupling, which can be tuned by $H$ and $x$, competes with quantum fluctuations from the transverse field, which result in a faster suppression of \Hc. The influence of the dilution of magnetic Ho$^{3+}$ ions by diamagnetic Y$^{3+}$ ions on the quantum phase transition could be quantitatively tracked by comparing the phase boundaries of experimental and mean-field phase diagrams.

\begin{acknowledgments}
We are indebted to Sebastian Gerischer, Bastian Klemke, and Klaus Kiefer for their help and support during the experiments, and especially for their expertise in designing and setting up a weak link for the dilution refrigerator. We are also grateful for insightful discussions with M.~Schechter during the preparation of this paper. We wish to thank the Swiss National Science Foundation (SNSF) for the financial support. I.K., J.O.P., and P.B. are grateful for the funding received from the ERC CONQUEST grant. This project has received funding from the European Union's Seventh Framework Programme for research, technological development and demonstration under the NMI3-II Grant No. 283883. Neutron scattering measurements were performed at SINQ, HZB, HML, and ILL facilities.
\end{acknowledgments}

\bibliography{shorttitles,biblio}

\begin{thebibliography}{39}%
\makeatletter
\providecommand \@ifxundefined [1]{%
 \@ifx{#1\undefined}
}%
\providecommand \@ifnum [1]{%
 \ifnum #1\expandafter \@firstoftwo
 \else \expandafter \@secondoftwo
 \fi
}%
\providecommand \@ifx [1]{%
 \ifx #1\expandafter \@firstoftwo
 \else \expandafter \@secondoftwo
 \fi
}%
\providecommand \natexlab [1]{#1}%
\providecommand \enquote  [1]{``#1''}%
\providecommand \bibnamefont  [1]{#1}%
\providecommand \bibfnamefont [1]{#1}%
\providecommand \citenamefont [1]{#1}%
\providecommand \href@noop [0]{\@secondoftwo}%
\providecommand \href [0]{\begingroup \@sanitize@url \@href}%
\providecommand \@href[1]{\@@startlink{#1}\@@href}%
\providecommand \@@href[1]{\endgroup#1\@@endlink}%
\providecommand \@sanitize@url [0]{\catcode `\\12\catcode `\$12\catcode
  `\&12\catcode `\#12\catcode `\^12\catcode `\_12\catcode `\%12\relax}%
\providecommand \@@startlink[1]{}%
\providecommand \@@endlink[0]{}%
\providecommand \url  [0]{\begingroup\@sanitize@url \@url }%
\providecommand \@url [1]{\endgroup\@href {#1}{\urlprefix }}%
\providecommand \urlprefix  [0]{URL }%
\providecommand \Eprint [0]{\href }%
\providecommand \doibase [0]{http://dx.doi.org/}%
\providecommand \selectlanguage [0]{\@gobble}%
\providecommand \bibinfo  [0]{\@secondoftwo}%
\providecommand \bibfield  [0]{\@secondoftwo}%
\providecommand \translation [1]{[#1]}%
\providecommand \BibitemOpen [0]{}%
\providecommand \bibitemStop [0]{}%
\providecommand \bibitemNoStop [0]{.\EOS\space}%
\providecommand \EOS [0]{\spacefactor3000\relax}%
\providecommand \BibitemShut  [1]{\csname bibitem#1\endcsname}%
\let\auto@bib@innerbib\@empty
\bibitem [{\citenamefont {Als-Nielsen}\ \emph {et~al.}(1974)\citenamefont
  {Als-Nielsen}, \citenamefont {Holmes},\ and\ \citenamefont
  {Guggenheim}}]{AlsNielsen1974}%
  \BibitemOpen
  \bibfield  {author} {\bibinfo {author} {\bibfnamefont {J.}~\bibnamefont
  {Als-Nielsen}}, \bibinfo {author} {\bibfnamefont {L.~M.}\ \bibnamefont
  {Holmes}}, \ and\ \bibinfo {author} {\bibfnamefont {H.~J.}\ \bibnamefont
  {Guggenheim}},\ }\href@noop {} {\bibfield  {journal} {\bibinfo  {journal}
  {Phys. Rev. Lett.}\ }\textbf {\bibinfo {volume} {32}},\ \bibinfo {pages}
  {610} (\bibinfo {year} {1974})}\BibitemShut {NoStop}%
\bibitem [{\citenamefont {Hansen}\ \emph {et~al.}(1975)\citenamefont {Hansen},
  \citenamefont {Johansson},\ and\ \citenamefont {Nevald}}]{Hansen1975}%
  \BibitemOpen
  \bibfield  {author} {\bibinfo {author} {\bibfnamefont {P.}~\bibnamefont
  {Hansen}}, \bibinfo {author} {\bibfnamefont {T.}~\bibnamefont {Johansson}}, \
  and\ \bibinfo {author} {\bibfnamefont {R.}~\bibnamefont {Nevald}},\
  }\href@noop {} {\bibfield  {journal} {\bibinfo  {journal} {Phys. Rev. B}\
  }\textbf {\bibinfo {volume} {12}},\ \bibinfo {pages} {5315} (\bibinfo {year}
  {1975})}\BibitemShut {NoStop}%
\bibitem [{\citenamefont {Misra}\ and\ \citenamefont
  {Felsteiner}(1977)}]{Misra1977}%
  \BibitemOpen
  \bibfield  {author} {\bibinfo {author} {\bibfnamefont {S.~K.}\ \bibnamefont
  {Misra}}\ and\ \bibinfo {author} {\bibfnamefont {J.}~\bibnamefont
  {Felsteiner}},\ }\href@noop {} {\bibfield  {journal} {\bibinfo  {journal}
  {Phys. Rev. B}\ }\textbf {\bibinfo {volume} {15}},\ \bibinfo {pages} {4309}
  (\bibinfo {year} {1977})}\BibitemShut {NoStop}%
\bibitem [{\citenamefont {Magari\~no}\ \emph {et~al.}(1980)\citenamefont
  {Magari\~no}, \citenamefont {Tuchendler}, \citenamefont {Beauvillain},\ and\
  \citenamefont {Laursen}}]{Magarino1980}%
  \BibitemOpen
  \bibfield  {author} {\bibinfo {author} {\bibfnamefont {J.}~\bibnamefont
  {Magari\~no}}, \bibinfo {author} {\bibfnamefont {J.}~\bibnamefont
  {Tuchendler}}, \bibinfo {author} {\bibfnamefont {P.}~\bibnamefont
  {Beauvillain}}, \ and\ \bibinfo {author} {\bibfnamefont {I.}~\bibnamefont
  {Laursen}},\ }\href@noop {} {\bibfield  {journal} {\bibinfo  {journal} {Phys.
  Rev. B}\ }\textbf {\bibinfo {volume} {21}},\ \bibinfo {pages} {18} (\bibinfo
  {year} {1980})}\BibitemShut {NoStop}%
\bibitem [{\citenamefont {Mennenga}\ \emph {et~al.}(1984)\citenamefont
  {Mennenga}, \citenamefont {de~Jongh}, \citenamefont {Huiskamp},\ and\
  \citenamefont {Laursen}}]{Mennenga1984}%
  \BibitemOpen
  \bibfield  {author} {\bibinfo {author} {\bibfnamefont {G.}~\bibnamefont
  {Mennenga}}, \bibinfo {author} {\bibfnamefont {L.~J.}\ \bibnamefont
  {de~Jongh}}, \bibinfo {author} {\bibfnamefont {W.~J.}\ \bibnamefont
  {Huiskamp}}, \ and\ \bibinfo {author} {\bibfnamefont {I.}~\bibnamefont
  {Laursen}},\ }\href@noop {} {\bibfield  {journal} {\bibinfo  {journal} {J.
  Mag. Mag. Mat.}\ }\textbf {\bibinfo {volume} {44}},\ \bibinfo {pages} {48}
  (\bibinfo {year} {1984})}\BibitemShut {NoStop}%
\bibitem [{\citenamefont {Kraemer}\ \emph {et~al.}(2012)\citenamefont
  {Kraemer}, \citenamefont {Nikseresht}, \citenamefont {Piatek}, \citenamefont
  {Tsyrulin}, \citenamefont {Dalla~Piazza}, \citenamefont {Kiefer},
  \citenamefont {Klemke}, \citenamefont {Rosenbaum}, \citenamefont {Aeppli},
  \citenamefont {Gannarelli} \emph {et~al.}}]{Kraemer2012}%
  \BibitemOpen
  \bibfield  {author} {\bibinfo {author} {\bibfnamefont {C.}~\bibnamefont
  {Kraemer}}, \bibinfo {author} {\bibfnamefont {N.}~\bibnamefont {Nikseresht}},
  \bibinfo {author} {\bibfnamefont {J.~O.}\ \bibnamefont {Piatek}}, \bibinfo
  {author} {\bibfnamefont {N.}~\bibnamefont {Tsyrulin}}, \bibinfo {author}
  {\bibfnamefont {B.}~\bibnamefont {Dalla~Piazza}}, \bibinfo {author}
  {\bibfnamefont {K.}~\bibnamefont {Kiefer}}, \bibinfo {author} {\bibfnamefont
  {B.}~\bibnamefont {Klemke}}, \bibinfo {author} {\bibfnamefont {T.~F.}\
  \bibnamefont {Rosenbaum}}, \bibinfo {author} {\bibfnamefont {G.}~\bibnamefont
  {Aeppli}}, \bibinfo {author} {\bibfnamefont {C.}~\bibnamefont {Gannarelli}},
  \emph {et~al.},\ }\href@noop {} {\bibfield  {journal} {\bibinfo  {journal}
  {Science}\ }\textbf {\bibinfo {volume} {336}},\ \bibinfo {pages} {1416}
  (\bibinfo {year} {2012})}\BibitemShut {NoStop}%
\bibitem [{\citenamefont {Babkevich}\ \emph {et~al.}(2016)\citenamefont
  {Babkevich}, \citenamefont {Jeong}, \citenamefont {Matsumoto}, \citenamefont
  {Kovacevic}, \citenamefont {Finco}, \citenamefont {Toft-Petersen},
  \citenamefont {Ritter}, \citenamefont {M\aa{}nsson}, \citenamefont
  {Nakatsuji},\ and\ \citenamefont {R\o{}nnow}}]{babkevich-prl-2016}%
  \BibitemOpen
  \bibfield  {author} {\bibinfo {author} {\bibfnamefont {P.}~\bibnamefont
  {Babkevich}}, \bibinfo {author} {\bibfnamefont {M.}~\bibnamefont {Jeong}},
  \bibinfo {author} {\bibfnamefont {Y.}~\bibnamefont {Matsumoto}}, \bibinfo
  {author} {\bibfnamefont {I.}~\bibnamefont {Kovacevic}}, \bibinfo {author}
  {\bibfnamefont {A.}~\bibnamefont {Finco}}, \bibinfo {author} {\bibfnamefont
  {R.}~\bibnamefont {Toft-Petersen}}, \bibinfo {author} {\bibfnamefont
  {C.}~\bibnamefont {Ritter}}, \bibinfo {author} {\bibfnamefont
  {M.}~\bibnamefont {M\aa{}nsson}}, \bibinfo {author} {\bibfnamefont
  {S.}~\bibnamefont {Nakatsuji}}, \ and\ \bibinfo {author} {\bibfnamefont
  {H.~M.}\ \bibnamefont {R\o{}nnow}},\ }\href {\doibase
  10.1103/PhysRevLett.116.197202} {\bibfield  {journal} {\bibinfo  {journal}
  {Phys. Rev. Lett.}\ }\textbf {\bibinfo {volume} {116}},\ \bibinfo {pages}
  {197202} (\bibinfo {year} {2016})}\BibitemShut {NoStop}%
\bibitem [{\citenamefont {Beauvillain}\ \emph {et~al.}(1978)\citenamefont
  {Beauvillain}, \citenamefont {Renard}, \citenamefont {Laursen},\ and\
  \citenamefont {Walker}}]{Beauvillain1978}%
  \BibitemOpen
  \bibfield  {author} {\bibinfo {author} {\bibfnamefont {P.}~\bibnamefont
  {Beauvillain}}, \bibinfo {author} {\bibfnamefont {J.~P.}\ \bibnamefont
  {Renard}}, \bibinfo {author} {\bibfnamefont {I.}~\bibnamefont {Laursen}}, \
  and\ \bibinfo {author} {\bibfnamefont {P.~J.}\ \bibnamefont {Walker}},\
  }\href@noop {} {\bibfield  {journal} {\bibinfo  {journal} {Phys. Rev. B}\
  }\textbf {\bibinfo {volume} {18}},\ \bibinfo {pages} {3360} (\bibinfo {year}
  {1978})}\BibitemShut {NoStop}%
\bibitem [{\citenamefont {Christensen}(1979)}]{Christensen1979}%
  \BibitemOpen
  \bibfield  {author} {\bibinfo {author} {\bibfnamefont {H.~P.}\ \bibnamefont
  {Christensen}},\ }\href@noop {} {\bibfield  {journal} {\bibinfo  {journal}
  {Phys. Rev. B}\ }\textbf {\bibinfo {volume} {19}},\ \bibinfo {pages} {6564}
  (\bibinfo {year} {1979})}\BibitemShut {NoStop}%
\bibitem [{\citenamefont {Chakraborty}\ \emph {et~al.}(2004)\citenamefont
  {Chakraborty}, \citenamefont {Henelius}, \citenamefont {Kj{\o}nsberg},
  \citenamefont {Sandvik},\ and\ \citenamefont {Girvin}}]{Chakraborty2004}%
  \BibitemOpen
  \bibfield  {author} {\bibinfo {author} {\bibfnamefont {P.~B.}\ \bibnamefont
  {Chakraborty}}, \bibinfo {author} {\bibfnamefont {P.}~\bibnamefont
  {Henelius}}, \bibinfo {author} {\bibfnamefont {H.}~\bibnamefont
  {Kj{\o}nsberg}}, \bibinfo {author} {\bibfnamefont {A.~W.}\ \bibnamefont
  {Sandvik}}, \ and\ \bibinfo {author} {\bibfnamefont {S.~M.}\ \bibnamefont
  {Girvin}},\ }\href@noop {} {\bibfield  {journal} {\bibinfo  {journal} {Phys.
  Rev. B}\ }\textbf {\bibinfo {volume} {70}},\ \bibinfo {pages} {144411}
  (\bibinfo {year} {2004})}\BibitemShut {NoStop}%
\bibitem [{\citenamefont {R\o{}nnow}\ \emph {et~al.}(2007)\citenamefont
  {R\o{}nnow}, \citenamefont {Jensen}, \citenamefont {Parthasarathy},
  \citenamefont {Aeppli}, \citenamefont {Rosenbaum}, \citenamefont {McMorrow},\
  and\ \citenamefont {Kraemer}}]{Ronnow2007}%
  \BibitemOpen
  \bibfield  {author} {\bibinfo {author} {\bibfnamefont {H.~M.}\ \bibnamefont
  {R\o{}nnow}}, \bibinfo {author} {\bibfnamefont {J.}~\bibnamefont {Jensen}},
  \bibinfo {author} {\bibfnamefont {R.}~\bibnamefont {Parthasarathy}}, \bibinfo
  {author} {\bibfnamefont {G.}~\bibnamefont {Aeppli}}, \bibinfo {author}
  {\bibfnamefont {T.~F.}\ \bibnamefont {Rosenbaum}}, \bibinfo {author}
  {\bibfnamefont {D.~F.}\ \bibnamefont {McMorrow}}, \ and\ \bibinfo {author}
  {\bibfnamefont {C.}~\bibnamefont {Kraemer}},\ }\href@noop {} {\bibfield
  {journal} {\bibinfo  {journal} {Phys. Rev. B}\ }\textbf {\bibinfo {volume}
  {75}},\ \bibinfo {pages} {054426} (\bibinfo {year} {2007})}\BibitemShut
  {NoStop}%
\bibitem [{\citenamefont {Schechter}\ and\ \citenamefont
  {Stamp}(2008)}]{Schechter2008}%
  \BibitemOpen
  \bibfield  {author} {\bibinfo {author} {\bibfnamefont {M.}~\bibnamefont
  {Schechter}}\ and\ \bibinfo {author} {\bibfnamefont {P.~C.~E.}\ \bibnamefont
  {Stamp}},\ }\href@noop {} {\bibfield  {journal} {\bibinfo  {journal} {Phys.
  Rev. B}\ }\textbf {\bibinfo {volume} {78}},\ \bibinfo {pages} {054438}
  (\bibinfo {year} {2008})}\BibitemShut {NoStop}%
\bibitem [{\citenamefont {R\o{}nnow}\ \emph {et~al.}(2005)\citenamefont
  {R\o{}nnow}, \citenamefont {Parthasarathy}, \citenamefont {Jensen},
  \citenamefont {Aeppli}, \citenamefont {Rosenbaum},\ and\ \citenamefont
  {McMorrow}}]{Ronnow2005}%
  \BibitemOpen
  \bibfield  {author} {\bibinfo {author} {\bibfnamefont {H.~M.}\ \bibnamefont
  {R\o{}nnow}}, \bibinfo {author} {\bibfnamefont {R.}~\bibnamefont
  {Parthasarathy}}, \bibinfo {author} {\bibfnamefont {J.}~\bibnamefont
  {Jensen}}, \bibinfo {author} {\bibfnamefont {G.}~\bibnamefont {Aeppli}},
  \bibinfo {author} {\bibfnamefont {T.~F.}\ \bibnamefont {Rosenbaum}}, \ and\
  \bibinfo {author} {\bibfnamefont {D.~F.}\ \bibnamefont {McMorrow}},\
  }\href@noop {} {\bibfield  {journal} {\bibinfo  {journal} {Science}\ }\textbf
  {\bibinfo {volume} {308}},\ \bibinfo {pages} {389} (\bibinfo {year}
  {2005})}\BibitemShut {NoStop}%
\bibitem [{\citenamefont {Silevitch}\ \emph {et~al.}(2010)\citenamefont
  {Silevitch}, \citenamefont {Aeppli},\ and\ \citenamefont
  {Rosenbaum}}]{Silevitch2010}%
  \BibitemOpen
  \bibfield  {author} {\bibinfo {author} {\bibfnamefont {D.~M.}\ \bibnamefont
  {Silevitch}}, \bibinfo {author} {\bibfnamefont {G.}~\bibnamefont {Aeppli}}, \
  and\ \bibinfo {author} {\bibfnamefont {T.~F.}\ \bibnamefont {Rosenbaum}},\
  }\href@noop {} {\bibfield  {journal} {\bibinfo  {journal} {Proc. Nat. Acad.
  Sci. USA}\ }\textbf {\bibinfo {volume} {107}},\ \bibinfo {pages} {2797}
  (\bibinfo {year} {2010})}\BibitemShut {NoStop}%
\bibitem [{\citenamefont {Reich}\ \emph {et~al.}(1990)\citenamefont {Reich},
  \citenamefont {Ellman}, \citenamefont {Yang}, \citenamefont {Rosenbaum},
  \citenamefont {Aeppli},\ and\ \citenamefont {Belanger}}]{Reich1990}%
  \BibitemOpen
  \bibfield  {author} {\bibinfo {author} {\bibfnamefont {D.~H.}\ \bibnamefont
  {Reich}}, \bibinfo {author} {\bibfnamefont {B.}~\bibnamefont {Ellman}},
  \bibinfo {author} {\bibfnamefont {J.}~\bibnamefont {Yang}}, \bibinfo {author}
  {\bibfnamefont {T.~F.}\ \bibnamefont {Rosenbaum}}, \bibinfo {author}
  {\bibfnamefont {G.}~\bibnamefont {Aeppli}}, \ and\ \bibinfo {author}
  {\bibfnamefont {D.~P.}\ \bibnamefont {Belanger}},\ }\href@noop {} {\bibfield
  {journal} {\bibinfo  {journal} {Phys. Rev. B}\ }\textbf {\bibinfo {volume}
  {42}},\ \bibinfo {pages} {4631} (\bibinfo {year} {1990})}\BibitemShut
  {NoStop}%
\bibitem [{\citenamefont {Tabei}\ \emph {et~al.}(2006)\citenamefont {Tabei},
  \citenamefont {Gingras}, \citenamefont {Kao}, \citenamefont {Stasiak},\ and\
  \citenamefont {Fortin}}]{Tabei2006}%
  \BibitemOpen
  \bibfield  {author} {\bibinfo {author} {\bibfnamefont {S.~M.~A.}\
  \bibnamefont {Tabei}}, \bibinfo {author} {\bibfnamefont {M.~J.~P.}\
  \bibnamefont {Gingras}}, \bibinfo {author} {\bibfnamefont {Y.~J.}\
  \bibnamefont {Kao}}, \bibinfo {author} {\bibfnamefont {P.}~\bibnamefont
  {Stasiak}}, \ and\ \bibinfo {author} {\bibfnamefont {J.~Y.}\ \bibnamefont
  {Fortin}},\ }\href@noop {} {\bibfield  {journal} {\bibinfo  {journal} {Phys.
  Rev. Lett.}\ }\textbf {\bibinfo {volume} {97}},\ \bibinfo {pages} {237203}
  (\bibinfo {year} {2006})}\BibitemShut {NoStop}%
\bibitem [{\citenamefont {Alonso}\ and\ \citenamefont
  {Fern\'andez}(2010)}]{Alonso2010}%
  \BibitemOpen
  \bibfield  {author} {\bibinfo {author} {\bibfnamefont {J.~J.}\ \bibnamefont
  {Alonso}}\ and\ \bibinfo {author} {\bibfnamefont {J.~F.}\ \bibnamefont
  {Fern\'andez}},\ }\href@noop {} {\bibfield  {journal} {\bibinfo  {journal}
  {Phys. Rev. B}\ }\textbf {\bibinfo {volume} {81}},\ \bibinfo {pages} {064408}
  (\bibinfo {year} {2010})}\BibitemShut {NoStop}%
\bibitem [{\citenamefont {Wu}\ \emph {et~al.}(1993)\citenamefont {Wu},
  \citenamefont {Bitko}, \citenamefont {Rosenbaum},\ and\ \citenamefont
  {Aeppli}}]{Wu1993}%
  \BibitemOpen
  \bibfield  {author} {\bibinfo {author} {\bibfnamefont {W.}~\bibnamefont
  {Wu}}, \bibinfo {author} {\bibfnamefont {D.}~\bibnamefont {Bitko}}, \bibinfo
  {author} {\bibfnamefont {T.~F.}\ \bibnamefont {Rosenbaum}}, \ and\ \bibinfo
  {author} {\bibfnamefont {G.}~\bibnamefont {Aeppli}},\ }\href@noop {}
  {\bibfield  {journal} {\bibinfo  {journal} {Phys. Rev. Lett.}\ }\textbf
  {\bibinfo {volume} {71}},\ \bibinfo {pages} {1919} (\bibinfo {year}
  {1993})}\BibitemShut {NoStop}%
\bibitem [{\citenamefont {Tam}\ and\ \citenamefont {Gingras}(2009)}]{Tam2009}%
  \BibitemOpen
  \bibfield  {author} {\bibinfo {author} {\bibfnamefont {K.~M.}\ \bibnamefont
  {Tam}}\ and\ \bibinfo {author} {\bibfnamefont {M.~J.~P.}\ \bibnamefont
  {Gingras}},\ }\href@noop {} {\bibfield  {journal} {\bibinfo  {journal} {Phys.
  Rev. Lett.}\ }\textbf {\bibinfo {volume} {103}},\ \bibinfo {pages} {087202}
  (\bibinfo {year} {2009})}\BibitemShut {NoStop}%
\bibitem [{\citenamefont {Rosenbaum}(1999)}]{Rosenbaum1999}%
  \BibitemOpen
  \bibfield  {author} {\bibinfo {author} {\bibfnamefont {T.~F.}\ \bibnamefont
  {Rosenbaum}},\ }\href@noop {} {\bibfield  {journal} {\bibinfo  {journal} {J.
  Phys.: Condens. Matter}\ }\textbf {\bibinfo {volume} {8}},\ \bibinfo {pages}
  {9759} (\bibinfo {year} {1999})}\BibitemShut {NoStop}%
\bibitem [{\citenamefont {Biltmo}\ and\ \citenamefont
  {Henelius}(2007)}]{Biltmo2007}%
  \BibitemOpen
  \bibfield  {author} {\bibinfo {author} {\bibfnamefont {A.}~\bibnamefont
  {Biltmo}}\ and\ \bibinfo {author} {\bibfnamefont {P.}~\bibnamefont
  {Henelius}},\ }\href@noop {} {\bibfield  {journal} {\bibinfo  {journal}
  {Phys. Rev. B}\ }\textbf {\bibinfo {volume} {76}},\ \bibinfo {pages} {054423}
  (\bibinfo {year} {2007})}\BibitemShut {NoStop}%
\bibitem [{\citenamefont {Biltmo}\ and\ \citenamefont
  {Henelius}(2008)}]{Biltmo2008}%
  \BibitemOpen
  \bibfield  {author} {\bibinfo {author} {\bibfnamefont {A.}~\bibnamefont
  {Biltmo}}\ and\ \bibinfo {author} {\bibfnamefont {P.}~\bibnamefont
  {Henelius}},\ }\href@noop {} {\bibfield  {journal} {\bibinfo  {journal}
  {Phys. Rev. B}\ }\textbf {\bibinfo {volume} {78}},\ \bibinfo {pages} {054437}
  (\bibinfo {year} {2008})}\BibitemShut {NoStop}%
\bibitem [{\citenamefont {Schechter}\ \emph {et~al.}(2007)\citenamefont
  {Schechter}, \citenamefont {Stamp},\ and\ \citenamefont
  {Laflorencie}}]{Schechter2007}%
  \BibitemOpen
  \bibfield  {author} {\bibinfo {author} {\bibfnamefont {M.}~\bibnamefont
  {Schechter}}, \bibinfo {author} {\bibfnamefont {P.~C.~E.}\ \bibnamefont
  {Stamp}}, \ and\ \bibinfo {author} {\bibfnamefont {N.}~\bibnamefont
  {Laflorencie}},\ }\href@noop {} {\bibfield  {journal} {\bibinfo  {journal}
  {J. Phys.: Condens. Matter}\ }\textbf {\bibinfo {volume} {19}},\ \bibinfo
  {pages} {145218} (\bibinfo {year} {2007})}\BibitemShut {NoStop}%
\bibitem [{\citenamefont {Ghosh}\ \emph {et~al.}(2002)\citenamefont {Ghosh},
  \citenamefont {Parthasarathy}, \citenamefont {Rosenbaum},\ and\ \citenamefont
  {Aeppli}}]{Ghosh2002}%
  \BibitemOpen
  \bibfield  {author} {\bibinfo {author} {\bibfnamefont {S.}~\bibnamefont
  {Ghosh}}, \bibinfo {author} {\bibfnamefont {R.}~\bibnamefont
  {Parthasarathy}}, \bibinfo {author} {\bibfnamefont {T.~F.}\ \bibnamefont
  {Rosenbaum}}, \ and\ \bibinfo {author} {\bibfnamefont {G.}~\bibnamefont
  {Aeppli}},\ }\href@noop {} {\bibfield  {journal} {\bibinfo  {journal}
  {Science}\ }\textbf {\bibinfo {volume} {296}},\ \bibinfo {pages} {2195}
  (\bibinfo {year} {2002})}\BibitemShut {NoStop}%
\bibitem [{\citenamefont {Reich}\ \emph {et~al.}(1987)\citenamefont {Reich},
  \citenamefont {Rosenbaum},\ and\ \citenamefont {Aeppli}}]{Reich1987}%
  \BibitemOpen
  \bibfield  {author} {\bibinfo {author} {\bibfnamefont {D.~H.}\ \bibnamefont
  {Reich}}, \bibinfo {author} {\bibfnamefont {T.~F.}\ \bibnamefont
  {Rosenbaum}}, \ and\ \bibinfo {author} {\bibfnamefont {G.}~\bibnamefont
  {Aeppli}},\ }\href@noop {} {\bibfield  {journal} {\bibinfo  {journal} {Phys.
  Rev. Lett.}\ }\textbf {\bibinfo {volume} {59}},\ \bibinfo {pages} {1969}
  (\bibinfo {year} {1987})}\BibitemShut {NoStop}%
\bibitem [{\citenamefont {Silevitch}\ \emph {et~al.}(2007)\citenamefont
  {Silevitch}, \citenamefont {Bitko}, \citenamefont {Brooke}, \citenamefont
  {Ghosh}, \citenamefont {Aeppli},\ and\ \citenamefont
  {Rosenbaum}}]{Silevitch2007}%
  \BibitemOpen
  \bibfield  {author} {\bibinfo {author} {\bibfnamefont {D.~M.}\ \bibnamefont
  {Silevitch}}, \bibinfo {author} {\bibfnamefont {D.}~\bibnamefont {Bitko}},
  \bibinfo {author} {\bibfnamefont {J.}~\bibnamefont {Brooke}}, \bibinfo
  {author} {\bibfnamefont {S.}~\bibnamefont {Ghosh}}, \bibinfo {author}
  {\bibfnamefont {G.}~\bibnamefont {Aeppli}}, \ and\ \bibinfo {author}
  {\bibfnamefont {T.~F.}\ \bibnamefont {Rosenbaum}},\ }\href@noop {} {\bibfield
   {journal} {\bibinfo  {journal} {Nature (London)}\ }\textbf {\bibinfo
  {volume} {448}},\ \bibinfo {pages} {567} (\bibinfo {year}
  {2007})}\BibitemShut {NoStop}%
\bibitem [{\citenamefont {Piatek}\ \emph {et~al.}(2013)\citenamefont {Piatek},
  \citenamefont {Dalla~Piazza}, \citenamefont {Nikseresht}, \citenamefont
  {Tsyrulin}, \citenamefont {\ifmmode \check{Z}\else
  \v{Z}\fi{}ivkovi\ifmmode~\acute{c}\else \'{c}\fi{}}, \citenamefont
  {Kr\"amer}, \citenamefont {Laver}, \citenamefont {Prokes}, \citenamefont
  {Mata\ifmmode~\check{s}\else \v{s}\fi{}}, \citenamefont {Christensen},\ and\
  \citenamefont {R\o{}nnow}}]{Piatek2013}%
  \BibitemOpen
  \bibfield  {author} {\bibinfo {author} {\bibfnamefont {J.~O.}\ \bibnamefont
  {Piatek}}, \bibinfo {author} {\bibfnamefont {B.}~\bibnamefont
  {Dalla~Piazza}}, \bibinfo {author} {\bibfnamefont {N.}~\bibnamefont
  {Nikseresht}}, \bibinfo {author} {\bibfnamefont {N.}~\bibnamefont
  {Tsyrulin}}, \bibinfo {author} {\bibfnamefont {I.}~\bibnamefont {\ifmmode
  \check{Z}\else \v{Z}\fi{}ivkovi\ifmmode~\acute{c}\else \'{c}\fi{}}}, \bibinfo
  {author} {\bibfnamefont {K.~W.}\ \bibnamefont {Kr\"amer}}, \bibinfo {author}
  {\bibfnamefont {M.}~\bibnamefont {Laver}}, \bibinfo {author} {\bibfnamefont
  {K.}~\bibnamefont {Prokes}}, \bibinfo {author} {\bibfnamefont
  {S.}~\bibnamefont {Mata\ifmmode~\check{s}\else \v{s}\fi{}}}, \bibinfo
  {author} {\bibfnamefont {N.~B.}\ \bibnamefont {Christensen}}, \ and\ \bibinfo
  {author} {\bibfnamefont {H.~M.}\ \bibnamefont {R\o{}nnow}},\ }\href@noop {}
  {\bibfield  {journal} {\bibinfo  {journal} {Phys. Rev. B}\ }\textbf {\bibinfo
  {volume} {88}},\ \bibinfo {pages} {014408} (\bibinfo {year}
  {2013})}\BibitemShut {NoStop}%
\bibitem [{\citenamefont {Kjaer}\ \emph {et~al.}(1999)\citenamefont {Kjaer},
  \citenamefont {Als-Nielsen}, \citenamefont {Laursen},\ and\ \citenamefont
  {Larsen}}]{Kjaer1999}%
  \BibitemOpen
  \bibfield  {author} {\bibinfo {author} {\bibfnamefont {K.}~\bibnamefont
  {Kjaer}}, \bibinfo {author} {\bibfnamefont {J.}~\bibnamefont {Als-Nielsen}},
  \bibinfo {author} {\bibfnamefont {I.}~\bibnamefont {Laursen}}, \ and\
  \bibinfo {author} {\bibfnamefont {F.~K.}\ \bibnamefont {Larsen}},\
  }\href@noop {} {\bibfield  {journal} {\bibinfo  {journal} {J. Phys.: Condens.
  Matter}\ }\textbf {\bibinfo {volume} {1}},\ \bibinfo {pages} {5743} (\bibinfo
  {year} {1999})}\BibitemShut {NoStop}%
\bibitem [{\citenamefont {Kovacevic}\ \emph {et~al.}()\citenamefont
  {Kovacevic}, \citenamefont {Babkevich}, \citenamefont {Jeong}, \citenamefont
  {Piatek}, \citenamefont {Boero},\ and\ \citenamefont
  {R{\o}nnow}}]{kovacevic-submitted}%
  \BibitemOpen
  \bibfield  {author} {\bibinfo {author} {\bibfnamefont {I.}~\bibnamefont
  {Kovacevic}}, \bibinfo {author} {\bibfnamefont {P.}~\bibnamefont
  {Babkevich}}, \bibinfo {author} {\bibfnamefont {M.}~\bibnamefont {Jeong}},
  \bibinfo {author} {\bibfnamefont {J.~O.}\ \bibnamefont {Piatek}}, \bibinfo
  {author} {\bibfnamefont {G.}~\bibnamefont {Boero}}, \ and\ \bibinfo {author}
  {\bibfnamefont {H.~M.}\ \bibnamefont {R{\o}nnow}},\ }\href@noop {} {\bibinfo
  {journal} {arXiv:1607.00124}\ }\BibitemShut {NoStop}%
\bibitem [{\citenamefont {Babkevich}\ \emph {et~al.}(2015)\citenamefont
  {Babkevich}, \citenamefont {Finco}, \citenamefont {Jeong}, \citenamefont
  {Dalla~Piazza}, \citenamefont {Kovacevic}, \citenamefont {Klughertz},
  \citenamefont {Kr\"amer}, \citenamefont {Kraemer}, \citenamefont {Adroja},
  \citenamefont {Goremychkin}, \citenamefont {Unruh}, \citenamefont
  {Str\"assle}, \citenamefont {Di~Lieto}, \citenamefont {Jensen},\ and\
  \citenamefont {R\o{}nnow}}]{babkevich-prb-2015}%
  \BibitemOpen
\bibfield  {journal} {  }\bibfield  {author} {\bibinfo {author} {\bibfnamefont
  {P.}~\bibnamefont {Babkevich}}, \bibinfo {author} {\bibfnamefont
  {A.}~\bibnamefont {Finco}}, \bibinfo {author} {\bibfnamefont
  {M.}~\bibnamefont {Jeong}}, \bibinfo {author} {\bibfnamefont
  {B.}~\bibnamefont {Dalla~Piazza}}, \bibinfo {author} {\bibfnamefont
  {I.}~\bibnamefont {Kovacevic}}, \bibinfo {author} {\bibfnamefont
  {G.}~\bibnamefont {Klughertz}}, \bibinfo {author} {\bibfnamefont {K.~W.}\
  \bibnamefont {Kr\"amer}}, \bibinfo {author} {\bibfnamefont {C.}~\bibnamefont
  {Kraemer}}, \bibinfo {author} {\bibfnamefont {D.~T.}\ \bibnamefont {Adroja}},
  \bibinfo {author} {\bibfnamefont {E.}~\bibnamefont {Goremychkin}}, \bibinfo
  {author} {\bibfnamefont {T.}~\bibnamefont {Unruh}}, \bibinfo {author}
  {\bibfnamefont {T.}~\bibnamefont {Str\"assle}}, \bibinfo {author}
  {\bibfnamefont {A.}~\bibnamefont {Di~Lieto}}, \bibinfo {author}
  {\bibfnamefont {J.}~\bibnamefont {Jensen}}, \ and\ \bibinfo {author}
  {\bibfnamefont {H.~M.}\ \bibnamefont {R\o{}nnow}},\ }\href {\doibase
  10.1103/PhysRevB.92.144422} {\bibfield  {journal} {\bibinfo  {journal} {Phys.
  Rev. B}\ }\textbf {\bibinfo {volume} {92}},\ \bibinfo {pages} {144422}
  (\bibinfo {year} {2015})}\BibitemShut {NoStop}%
\bibitem [{\citenamefont {Jensen}(1984)}]{jensen-jopc-1984}%
  \BibitemOpen
  \bibfield  {author} {\bibinfo {author} {\bibfnamefont {J.}~\bibnamefont
  {Jensen}},\ }\href {http://stacks.iop.org/0022-3719/17/i=30/a=011} {\bibfield
   {journal} {\bibinfo  {journal} {J. Phys. C}\ }\textbf {\bibinfo {volume}
  {17}},\ \bibinfo {pages} {5367} (\bibinfo {year} {1984})}\BibitemShut
  {NoStop}%
\bibitem [{\citenamefont {Tabei}\ \emph {et~al.}(2008)\citenamefont {Tabei},
  \citenamefont {Gingras}, \citenamefont {Kao},\ and\ \citenamefont
  {Yavors'kii}}]{Tabei2008}%
  \BibitemOpen
  \bibfield  {author} {\bibinfo {author} {\bibfnamefont {S.~M.~A.}\
  \bibnamefont {Tabei}}, \bibinfo {author} {\bibfnamefont {M.~J.~P.}\
  \bibnamefont {Gingras}}, \bibinfo {author} {\bibfnamefont {Y.~J.}\
  \bibnamefont {Kao}}, \ and\ \bibinfo {author} {\bibfnamefont
  {T.}~\bibnamefont {Yavors'kii}},\ }\href@noop {} {\bibfield  {journal}
  {\bibinfo  {journal} {Phys. Rev. B}\ }\textbf {\bibinfo {volume} {78}},\
  \bibinfo {pages} {184408} (\bibinfo {year} {2008})}\BibitemShut {NoStop}%
\bibitem [{\citenamefont {Dunn}\ \emph {et~al.}(2012)\citenamefont {Dunn},
  \citenamefont {Stahl}, \citenamefont {Macdonald}, \citenamefont {Liu},
  \citenamefont {Reshitnyk}, \citenamefont {Sim},\ and\ \citenamefont
  {Hill}}]{dunn-prb-2012}%
  \BibitemOpen
  \bibfield  {author} {\bibinfo {author} {\bibfnamefont {J.~L.}\ \bibnamefont
  {Dunn}}, \bibinfo {author} {\bibfnamefont {C.}~\bibnamefont {Stahl}},
  \bibinfo {author} {\bibfnamefont {A.~J.}\ \bibnamefont {Macdonald}}, \bibinfo
  {author} {\bibfnamefont {K.}~\bibnamefont {Liu}}, \bibinfo {author}
  {\bibfnamefont {Y.}~\bibnamefont {Reshitnyk}}, \bibinfo {author}
  {\bibfnamefont {W.}~\bibnamefont {Sim}}, \ and\ \bibinfo {author}
  {\bibfnamefont {R.~W.}\ \bibnamefont {Hill}},\ }\href {\doibase
  10.1103/PhysRevB.86.094428} {\bibfield  {journal} {\bibinfo  {journal} {Phys.
  Rev. B}\ }\textbf {\bibinfo {volume} {86}},\ \bibinfo {pages} {094428}
  (\bibinfo {year} {2012})}\BibitemShut {NoStop}%
\bibitem [{\citenamefont {Jensen}\ and\ \citenamefont
  {Mackintosh}(1991)}]{Jensen1991}%
  \BibitemOpen
  \bibfield  {author} {\bibinfo {author} {\bibfnamefont {J.}~\bibnamefont
  {Jensen}}\ and\ \bibinfo {author} {\bibfnamefont {A.~R.}\ \bibnamefont
  {Mackintosh}},\ }\href@noop {} {\emph {\bibinfo {title} {Rare Earth
  Magnetism: Structures and Excitations}}}\ (\bibinfo  {publisher} {Clarendon
  Press, Oxford, UK},\ \bibinfo {year} {1991})\BibitemShut {NoStop}%
\bibitem [{\citenamefont {Kraemer}(2009)}]{Kraemer2009}%
  \BibitemOpen
  \bibfield  {author} {\bibinfo {author} {\bibfnamefont {C.}~\bibnamefont
  {Kraemer}},\ }\emph {\bibinfo {title} {Quantum phase transitions in a
  magnetic model system}},\ \href@noop {} {Ph.D. thesis},\ \bibinfo  {school}
  {ETH} (\bibinfo {year} {2009})\BibitemShut {NoStop}%
\bibitem [{\citenamefont {Andresen}\ \emph {et~al.}(2013)\citenamefont
  {Andresen}, \citenamefont {Thomas}, \citenamefont {Katzgraber},\ and\
  \citenamefont {Schechter}}]{andresen-prl-2013}%
  \BibitemOpen
  \bibfield  {author} {\bibinfo {author} {\bibfnamefont {J.~C.}\ \bibnamefont
  {Andresen}}, \bibinfo {author} {\bibfnamefont {C.~K.}\ \bibnamefont
  {Thomas}}, \bibinfo {author} {\bibfnamefont {H.~G.}\ \bibnamefont
  {Katzgraber}}, \ and\ \bibinfo {author} {\bibfnamefont {M.}~\bibnamefont
  {Schechter}},\ }\href {\doibase 10.1103/PhysRevLett.111.177202} {\bibfield
  {journal} {\bibinfo  {journal} {Phys. Rev. Lett.}\ }\textbf {\bibinfo
  {volume} {111}},\ \bibinfo {pages} {177202} (\bibinfo {year}
  {2013})}\BibitemShut {NoStop}%
\bibitem [{\citenamefont {Bitko}\ \emph {et~al.}(1996)\citenamefont {Bitko},
  \citenamefont {Rosenbaum},\ and\ \citenamefont {Aeppli}}]{Bitko1996}%
  \BibitemOpen
  \bibfield  {author} {\bibinfo {author} {\bibfnamefont {D.}~\bibnamefont
  {Bitko}}, \bibinfo {author} {\bibfnamefont {T.~F.}\ \bibnamefont
  {Rosenbaum}}, \ and\ \bibinfo {author} {\bibfnamefont {G.}~\bibnamefont
  {Aeppli}},\ }\href@noop {} {\bibfield  {journal} {\bibinfo  {journal} {Phys.
  Rev. Lett.}\ }\textbf {\bibinfo {volume} {77}},\ \bibinfo {pages} {940}
  (\bibinfo {year} {1996})}\BibitemShut {NoStop}%
\bibitem [{\citenamefont {Schechter}\ and\ \citenamefont
  {Laflorencie}(2006)}]{sch2006}%
  \BibitemOpen
  \bibfield  {author} {\bibinfo {author} {\bibfnamefont {M.}~\bibnamefont
  {Schechter}}\ and\ \bibinfo {author} {\bibfnamefont {N.}~\bibnamefont
  {Laflorencie}},\ }\href@noop {} {\bibfield  {journal} {\bibinfo  {journal}
  {Phys. Rev. Lett.}\ }\textbf {\bibinfo {volume} {97}},\ \bibinfo {pages}
  {137204} (\bibinfo {year} {2006})}\BibitemShut {NoStop}%
\bibitem [{\citenamefont {Schechter}(2008)}]{schechter-prb-2008}%
  \BibitemOpen
  \bibfield  {author} {\bibinfo {author} {\bibfnamefont {M.}~\bibnamefont
  {Schechter}},\ }\href {\doibase 10.1103/PhysRevB.77.020401} {\bibfield
  {journal} {\bibinfo  {journal} {Phys. Rev. B}\ }\textbf {\bibinfo {volume}
  {77}},\ \bibinfo {pages} {020401} (\bibinfo {year} {2008})}\BibitemShut
  {NoStop}%
\end{thebibliography}%

\end{document}